\documentclass[twocolumn]{jpsj3} %% for short notes
\usepackage{graphicx}
\usepackage{bm}
\def\gsim{\mathop {\vtop {\ialign {##\crcr $\hfil \displaystyle {>}\hfil $\crcr \noalign {\kern1pt \nointerlineskip } $\,\sim$ \crcr \noalign {\kern1pt}}}}\limits}
\def\lsim{\mathop {\vtop {\ialign {##\crcr $\hfil \displaystyle {<}\hfil $\crcr \noalign {\kern1pt \nointerlineskip } $\,\,\sim$ \crcr \noalign {\kern1pt}}}}\limits}
\usepackage{color}

\title{Anomalous  $^{125}$Te Nuclear Spin Relaxation Coincident with Charge Kondo Behavior \\
in  Superconducting Pb$_{1-x}$Tl$_{x}$Te}

\author{
Hidekazu Mukuda$^{1}$\thanks{E-mail: mukuda@mp.es.osaka-u.ac.jp}, Takashi Matsumura$^{1}$, Shota Maki$^{1}$, Mitsuharu Yashima$^{1}$, Yoshio Kitaoka$^{1}$, \\
Kazumasa Miyake$^{2}$, Hironaru Murakami$^{3}$, Paula Giraldo-Gallo$^{4}$, 
Theodore  H. Geballe$^{4}$, and Ian R. Fisher$^{4}$
}

\inst{
$^{1}$Graduate School of Engineering Science, Osaka University, Toyonaka, Osaka 560-8531, Japan \\
$^{2}$Center for Advanced High Magnetic Field Science, Osaka University, Osaka 560-0043, Japan \\
$^{3}$Institute of Laser Engineering, Osaka University, Suita, Osaka 565-0871, Japan \\
$^{4}$Department of Applied Physics, Stanford University, California 94305, USA \\
}

\abst{
We report the results of a $^{125}$Te NMR study of single crystalline Pb$_{1-x}$Tl$_x$Te ($x$=0, 0.35, 1.0\%) as a window on the novel electronic states associated with the thallium impurities in PbTe.
The Knight shift is enhanced as $x$ increases, corresponding to an increase in the average density of states (DOS) coupled to a strong spatial variation in the local DOS surrounding each Tl dopant.  
Remarkably, for the superconducting composition ($x$=1.0\%), the $^{125}$Te nuclear spin relaxation rate ($1/T_1T$) for Te ions that are close to the Tl dopants is unexpectedly enhanced in the normal state below a characteristic temperature of $\sim$10 K,  below which the resistivity experiences an upturn. 
Such a simultaneous upturn in both the resistivity and $(1/T_1T)$ was not suppressed in the high magnetic field. 
We suggest that these observations are consistently accounted for by dynamical charge fluctuations in the absence of paramagnetism, which is anticipated by the charge Kondo scenario associated with the Tl dopants. 
In contrast, such anomalies were not detected in the non-superconducting samples ($x$=0 and 0.35\%),  suggesting a  connection between dynamical valence fluctuations and the occurrence of superconductivity in Pb$_{1-x}$Tl$_{x}$Te. 
}

\recdate{\today}
\newpage
\begin{document}
\maketitle

%\section{Introduction}===========================================================

%Search for unconventional superconducting mechanism is one of the possible routes to obtain higher temperature superconductor. 
Low-carrier-density superconductors provide a challenge to the conventional Bardeen-Cooper-Schrieffer (BCS) theory of superconductivity. Moreover, such materials potentially provide clues for ways to boost the critical temperature of other, more conventional, materials. 
Here, we investigate via NMR measurements the local electronic properties of one such anomalous low-carrier-density superconductor, Pb$_{1-x}$Tl$_{x}$Te.
PbTe is a narrow-gap semiconductor. 
Small amounts of substitution of Tl for Pb (i.e., Pb$_{1-x}$Tl$_{x}$Te)  lead to a superconducting (SC) ground state when $x$ exceeds $x_c\sim$0.3\% \cite{Chernik,Nemov,Murakami,Matsushita}. 
Significantly, thallium is the only dopant known to cause superconductivity in PbTe, suggesting that these specific impurities have a unique effect on the electronic states near the Fermi energy.
Although the carrier densities are $\lesssim$10$^{20}$cm$^{-3}$, the SC transition temperature rises to $T_c \sim$1.5 K for $x\sim$1.5\% (the solubility limit),  higher than that of other well-known low-carrier-density superconductors, such as SrTiO$_3$\cite{Schooley}.
%Although carrier densities are as low as $\sim$10$^{20}$cm$^{-3}$, the SC transition temperature $T_c \sim$1.5 K for $x\sim$1.5 \% (the solubility limit) is relatively higher than that of other well-known low-carrier density superconductors, such as SrTiO$_3$ for comparable carrier densities\cite{Schooley}. 
The hole density $p$, estimated by Hall coefficient measurements, increases linearly  with $x$ for compositions up to $x \sim x_c$, implying that each Tl impurity acts as an acceptor, having a formal valence  Tl$^{1+}$.
%exhibits a linear  with $x$, which is anticipated by consideration of the formal-valence substitution of Tl$^{1+}$ for Pb$^{2+}$, up to $x \sim x_c$. 
However, as $x$ increases further, the increase in $p$ gradually saturates, implying that impurities no longer contribute one hole per dopant.
%the additional carriers are partially compensated. 
Drawing on the known valence-skipping character of thallium ions \cite{Varma}, this behavior has been interpreted in terms of the onset of a degeneracy of impurity states with a formal valence of Tl$^{1+}$(hole doping) and Tl$^{3+}$(electron doping) for $x>x_c$ \cite{Nemov,Murakami,Matsushita,Matsushita2}. 
%It is relevant with an empirical fact that the Tl dopant is known to take either Tl$^{1+}$($6s^2$) or Tl$^{3+}$($6s^0$), but to skip an intermediate valence Tl$^{2+}$($6s^1$) due to the stability of a filled shell electron configuration. 
Indirect support for such a scenario was obtained via the observation of a logarithmic upturn in the resistivity at low temperatures for $x > x_c$, reminiscent of the Kondo effect\cite{Kondo}, but in the absence of unpaired spins\cite{Andronik,Matsushita}. 
This observation was interpreted as evidence for a \emph{charge} Kondo effect, that is, a Kondo effect arising from the interaction of the conduction electrons with the two degenerate valence states of the Tl dopants\cite{Matsushita,Dzero,Taraphder,Costi,Yanagisawa,Matsuura}. 
%, where pseudo-spin flip processes proceed via virtual excitations to the skipped valence state.
The fact that such an effect is observed only for SC compositions \cite{Matsushita,Erickson} implies that valence fluctuations might play a key role in the pairing interaction in Pb$_{1-x}$Tl$_{x}$Te, possibly explaining the high critical temperatures found in this system  \cite{Varma,Taraphder,Costi,Dzero,Yanagisawa,Matsuura}.
Motivated by these bulk experiments and theoretical insights, we investigated the \emph{local} electronic states around Tl dopants by means of $^{125}$Te-NMR from the microscopic point of view.

%, in particular using the Te ions to probe the response as a function of position relative to the Tl ions.  
%As a matter of fact, some degenerate state of Tl$^{1+}$ and Tl$^{3+}$ was corroborated by an observation of a logarithmic upturn in resistivity at low temperature for $x > x_c$ that reminds us of the spin-Kondo effect\cite{Andronik,Matsushita}.
%This is because two degenerate charge states of  2$e^-$(Tl$^{1+}$)  and 0$e^-$(Tl$^{3+}$) is possible to form a resonating valence state, which has been theoretically accounted for by {\it ``charge''} Kondo effect\cite{Varma,Taraphder,Costi,Dzero,Yanagisawa,Matsuura} in analogy with two degenerate spin states in  ``{\it spin}'' Kondo effect\cite{Kondo}.
%The fact that the  {\it ``charge''} Kondo effect is observed only in the SC samples for $x\ge x_c$\cite{Matsushita,Erickson} have attracted theoretical interests on an unconventional SC pairing mechanism  in terms of  ``{\it negative-U}'' model that introduces seemingly an {\it attractive} on-site interaction\cite{Varma,Taraphder,Costi,Dzero,Yanagisawa,Matsuura}.
%These experiments and theoretical works have motivated us to investigate local electronic states around valence skipping Tl dopants from microscopic points of view by means of NMR. 

In this Letter, we report  a systematic $^{125}$Te-NMR study on Pb$_{1-x}$Tl$_{x}$Te,  revealing  the unusual character of the local electronic states introduced by Tl dopants through systematic measurements of the Knight shift ($K$) and the nuclear spin relaxation rate $(1/T_1T)$. 
Our main result is that $(1/T_1T)$ is found to be significantly enhanced below 10 K at Te sites in the vicinity of the Tl dopants for the SC sample ($x$=1.0\%), whereas no such anomalous enhancement  is seen for $x$ = 0 (the non-SC parent compound) or $x\sim$ 0.35\% (i.e., $x \sim x_c$).  
The temperature at which $(1/T_1T)$ is enhanced for the SC  composition coincides with the temperature below which the normal state resistivity experiences an upturn. 
We suggest that such a simultaneous upturn in both the resistivity and $(1/T_1T)$ can be consistently understood in terms of the coherent charge fluctuations anticipated in the charge Kondo effect.

%In this Letter, we report $^{125}$Te-NMR study on Pb$_{1-x}$Tl$_{x}$Te, which unravels the local electronic states introduced by Tl dopants through the systematic measurements of Knight shift ($K$) and nuclear spin relaxation rate $(1/T_1T)$. 
%It is unexpected that $(1/T_1T)$ is significantly enhanced below 10 K at the Te sites in the vicinity of the Tl dopants for the SC sample ($x$=1.0 \%), but no anomaly has been seen for the non-SC sample ($x$=0.35 \%). 
%This NMR evidence demonstrates that the resonating valence state between 2$e^-$(Tl$^{1+}$)  and 0$e^-$(Tl$^{3+}$) results in the development of coherent $6s$-electron pair-hopping upon cooling below 10 K, increasing $(1/T_1T)$ around the Tl dopants for the SC sample. 
%This event may be in analogy with the increase in $(1/T_1T)$ just below $T_c$ due to the formation of Cooper-pairing, that is known as the coherence effect for BCS superconductors. 
%As a result, we propose that the coherent hopping of $6s$-electron pair may develop a Cooper-pairing formation to cause the superconductivity below $T_c\sim$1 K, as suggested by the negative-$U$ scenario\cite{Taraphder,Costi,Dzero,Yanagisawa,Matsuura}.

%\section{Experimental Procedures}================================================

\begin{figure}[tbp]
\centering
\includegraphics[width=0.95\linewidth]{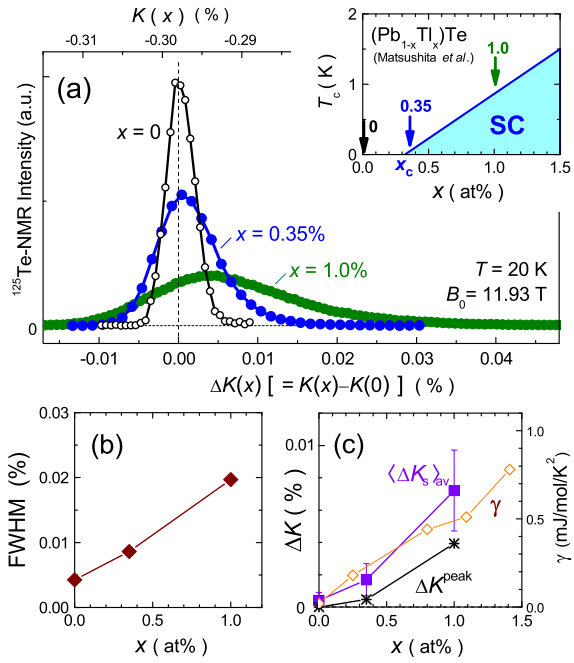}
\caption[]{(Color online) 
(a) $^{125}$Te-NMR spectra at $T$=20 K for $x$=0, 0.35, and 1.0\%.
Here, the horizontal axis  $\Delta K(x)$ is defined as $K(x)\!-\!K(0)$, which represents the relative shift from that of $x$=0. 
The arrows in the inset indicate the samples measured in this work, plotted on the SC phase diagram reported previously \cite{Matsushita}. 
%The $x$ dependence of (b) FWHM of the spectra represented in the scale of $K(x)$, and (c)  $\Delta K^{\rm peak}$ and $\langle \Delta K_s\rangle_{\rm av}$ defined by the peak and the center of gravity of the spectra, respectively. 
(b) The FWHM of the spectra, expressed as a scale of $K(x)$, and (c)  $\Delta K^{\rm peak}$ and $\langle \Delta K_s\rangle_{\rm av}$ defined by the peak and the center of gravity of the spectra, respectively, as a function of thallium concentration, $x$. 
The $x$ dependence of the Sommerfeld coefficient ($\gamma$) in specific heat is cited from Ref.\cite{Matsushita2}
}
\label{f1}
\end{figure}

High-quality single crystals of Pb$_{1-x}$Tl$_{x}$Te with $x$=0, 0.35, and 1.0\% were grown by an unseeded physical vapor transport method, as described previously\cite{Matsushita}. 
Thallium concentrations were determined for other crystals from the same growth batches by an electron microprobe analysis with an estimated uncertainty of $\sim\pm$ 0.1\% \cite{Matsushita}. 
The crystals of $x$=0.35 and 1.0\% correspond to the borderline region of the SC phase ($x\simeq x_c$)  and the SC region ($x > x_c$) with $T_c\sim$1 K, respectively, as shown by the arrows in the inset of Fig. \ref{f1}(a)\cite{Matsushita}. 
Resistivity measurements for these samples reproduce previous results,\cite{Matsushita} including the upturn  below $\sim$10 K for $x$=1\% (see inset of Fig. 3).
$^{125}$Te-NMR ($I$=1/2) measurements of the Knight shift ($K=(f-\gamma_nB_0)/\gamma_nB_0$) and the nuclear spin-lattice relaxation rate $(1/T_1)$ were performed on coarsely ground crystals of all three compositions  in a magnetic field of $B_0\sim$ 11.93 T. 
Here, $\gamma_n$ is the nuclear gyromagnetic ratio, and $f$ is the frequency. 
%\textcolor{blue}{ 
%We chose $^{125}$Te-nucleus for NMR probe since the Te is possible to occupy around the Tl-dopants that replace the Pb site. 
%}
$T_1$ was obtained by fitting a recovery curve for $^{125}$Te nuclear magnetization to a single exponential function, $m(t)\equiv [M_0-M(t)]/M_0=\exp \left(-t/T_1\right)$, where $M_0$ and $M(t)$ are the nuclear magnetizations for a thermal equilibrium condition and at time $t$ after a saturation pulse, respectively. 

%\section{results} ===========================================================
%fig1------------------------------------------------ 
%------------------------------------------------ 

First, we focus on the doping effect on the static electronic properties through the measurement of $K$.  
Figure \ref{f1}(a) shows the $^{125}$Te-NMR spectra at $T$= 20 K for $x$=0, 0.35, and 1.0\%.
Here, the frequency on the horizontal axis is converted to the scale of the Knight shift $K(x)$, and $\Delta K(x)$=$K(x)\!-\!K(0)$ represents the relative shift from that of undoped $x$=0 (bottom axis). 
The full-width at half-maximum (FWHM) of the spectra  becomes large with increasing $x$, as shown in Fig. \ref{f1}(b). 
The spectra are nevertheless composed of a single peak even for the doped samples, indicating that the doped Tl atoms are distributed randomly throughout the samples, excluding the possibility of dopant clusters and/or decomposition. 
The doping of the Tl atoms causes $\Delta K(x)$ to increase, as shown in Fig. \ref{f1}(c) by $\Delta K^{\rm peak}$ determined by the peak of spectra.
In general, the observed $K(x)$ comprises the spin shift $K_s$ and the chemical shift $K_{\rm chem}$. 
The former $K_s$  is proportional to $A_{\rm hf}\chi_0\propto\!A_{\rm hf}N_0$, where $\chi_0$ is the static spin susceptibility at ${q}$=0, $\!A_{\rm hf}$ is the hyperfine coupling constant, and $N_0$ is the density of states (DOS) at the Fermi level ($E_{\rm F}$).  
Since $K_{\rm chem}$ is  independent of $x$ in a small range,  $\Delta K(x)[\equiv K(x)\!-\!K(0)]$ corresponds to  their spin components $\Delta K_s(x) [\equiv K_s(x)\!-\!K_s(0)]$. 
Hence, in Fig. \ref{f1}(c), the increase in $\Delta K^{\rm peak}$ with Tl doping originates from that of $\chi_0$ or  $N_0$ at the Te sites in the vicinity of the Tl dopants. 
It is noteworthy that, as indicated in Fig. \ref{f1}(a), the FWHM in the spectrum is significantly increased toward the {\it positive} side in $\Delta K(x)$ with increasing $x$. 
The fraction of Te ions (relative to the total number of Te ions) that are nearest neighbors to a Tl dopant increases as $x$ increases; 
accordingly, the Te sites closer to the Tl dopants possess the larger $\Delta K_s(x)$ or the larger $N_0$ relative to those that are farther from the Tl dopants. 
In this context, the spatially averaged value of $\Delta K_s$ at the Te sites should be evaluated by $\langle \Delta K_s\rangle_{\rm av}$, which is defined by the center of gravity of the broad spectrum, as shown in Fig. \ref{f1}(c). 
The $x$ dependence of $\langle \Delta K_s\rangle_{\rm av}$  is comparable to that of the $N_0$ value  estimated by the Sommerfeld coefficient ($\gamma$) in specific heat measurements\cite{Matsushita2}. 
The close correspondence between these two spatially averaged values confirms our interpretation of the physical origin of $\Delta K(x)$. 
It also indicates that $K_{\rm chem}$ is assumed to be constant for the present $x$ region. 
We emphasize that the large distribution of $\Delta K_s(x)$ with increasing $x$ is indicative of a strong spatial variation in the local DOS surrounding each Tl dopant, which is especially pronounced for the larger Tl concentrations.

%fig2------------------------------------------------
\begin{figure}[tbp]
\centering
\includegraphics[width=1\linewidth]{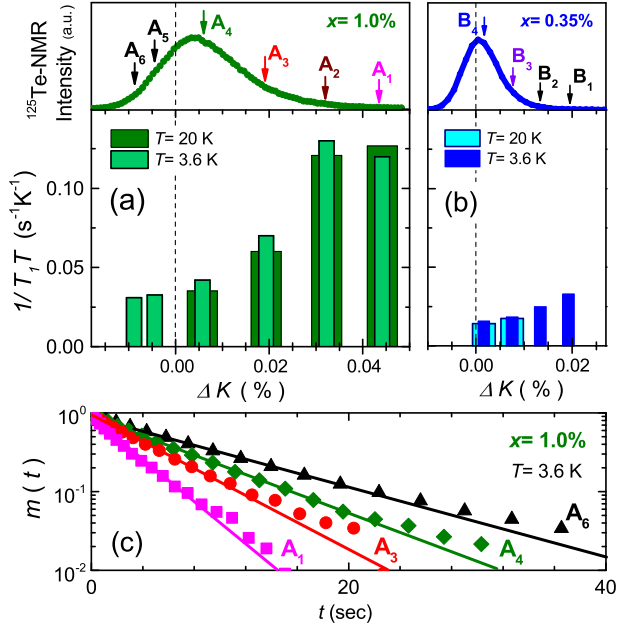}
\caption[]{(Color online) 
$(1/T_1T)$ for (a) $x$=1.0\% and (b) $x$=0.35\% measured as a function of $\Delta K$, corresponding to  A$_i$ ($i$=1 to 6) and B$_i$ ($i$=1 to 4) denoted in the upper panels, respectively.
%\cite{definition}. 
(c)  Recovery curves of nuclear magnetization $m(t)$ for  A$_{i}$ in $x$=1.0\%, the slope of which enables us to evaluate $T_1$.
Representative data are shown here for the sample with $x$=1.0\% at $T\sim$ 3.6 K. 
The large $(1/T_1T)$ value that is shown here for  the larger values of  $\Delta K$ is not observed in samples with $x$=0 and is therefore due to the presence of Tl dopants. 
%The large $(1/T_1T)$ at the large $\Delta K$ appears when $x\ne0$ due to the presence of Tl dopants, since such anomaly is not observed in $x$=0. 
}
\label{f2}
\end{figure}
%------------------------------------------------

Next, we address the evolution of the local electronic states introduced by the Tl dopants through the $^{125}$Te nuclear spin lattice relaxation rate, $1/T_1$. 
Figure \ref{f2}(a) shows  $(1/T_1T)$ for $x$=1.0\% at $T$=3.6 and 20 K, which are measured at the Te sites denoted by A$_i$ ($i$=1 to 6) in the spectrum (defined in the upper panel of the same figure).
%\cite{definition}. 
Each $T_1$ value for each A$_i$ is determined from the dominant component in the recovery curve $m(t)$, shown by solid lines in Fig. \ref{f2}(c). 
Note that the values of $\Delta K$ are widely distributed over the sample, which allows us to examine the  local electronic characteristics at the different distances from the Tl dopants.
In particular, we find that values of $(1/T_1T)$ become progressively larger for larger values of $\Delta K$.
%In fact, the values of $(1/T_1T)$ becomes large as $\Delta K$ increases. 
This is in sharp contrast to the undoped ($x$=0) case, which exhibits a homogeneous value of $1/T_1$ for all Te sites.  
It should be noted that the large values of $(1/T_1T)$ at the large values of $\Delta K$ originate from the Te sites in the vicinity of the Tl dopants, which is indicative of a significant change in the local electronic state induced by doping Tl atoms. 
In contrast, for $x$=0.35\% the distributions at the Te sites denoted by B$_i$ ($i$=1 to 4) are also significant  but smaller than those for $x$=1.0\%, as shown in Fig.~\ref{f2}(b). 
This shows that the presence of Tl dopants  increases the distributions in $(1/T_1T)$ and $\Delta K$ with increasing $x$($>$0).
%We remark that for $x$=1.0 \% with $T_c$=1.0 K, the $(1/T_1T)$ are markedly enhanced at the near neighbor Te sites to the Tl dopants.
%Clearly there is something markedly different about the local electronic structure associated with Tl dopants for the superconducting and non-superconducting compositions. 

%fig3------------------------------------------------
\begin{figure}[tbp]
\centering
\includegraphics[width=0.95\linewidth]{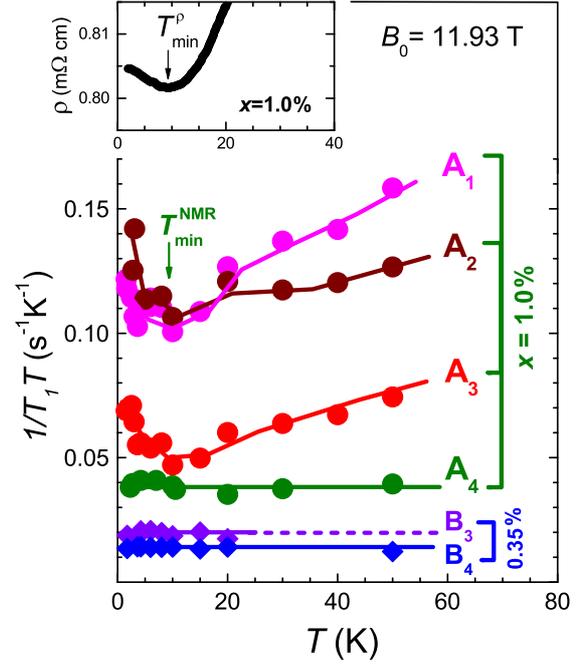}
\caption[]{(Color online)
$T$ dependence of  $(1/T_1T)$ for $x$=1.0 and 0.35\%  measured at $\Delta K$ denoted as  A$_{i}$  and B$_{i}$, respectively 
(see the upper panels of Figs. \ref{f2}(a) and \ref{f2}(b) for definitions of A$_{i}$  and B$_{i}$).
%\cite{definition}. 
The anomalous increases in  $(1/T_1T)$ below $T_{\rm min}^{\rm NMR}\sim$10 K for $x$=1.0\% are only seen for A$_i$($i$=1,2,3), corresponding to the Te sites close to the Tl dopants. 
Inset shows the Kondo-like resistivity upturn below $T_{\rm min}^\rho\sim$10 K observed in the present ($x$ = 1.0\%) sample\cite{Matsushita}.
}
\label{f3}
\end{figure}
%------------------------------------------------

In order to gain further insight into the unique electronic state induced by Tl dopants, and the difference between SC and non-SC compositions, the temperature ($T$) dependence of $(1/T_1T)$ was measured as shown in Fig.~\ref{f3}. 
For $x$=0.35\% ($x\simeq x_c$), $(1/T_1T)$=$const.$ is observed at B$_4$ and B$_3$ in the $T$ range of 1.4$-$60 K, as anticipated for a typical metal. 
In sharp contrast, for $x$=1.0\% ($x > x_c$), the $(1/T_1T)$ values at  A$_1$, A$_2$, and A$_3$ (i.e., in the vicinity of the Tl dopants) start to increase dramatically below $T_{\rm min}^{\rm NMR}\simeq$10 K.  
Significantly, $T_{\rm min}^{\rm NMR}\simeq$10 K coincides with $T_{\rm min}^\rho$, the temperature below which the resistivity experiences a logarithmic upturn \cite{Matsushita}, as shown in the inset of  Fig.~\ref{f3}. 
By contrast, such an increase in $(1/T_1T)$ at  A$_i$ ($i$=1-3)  is largely  suppressed at A$_4$, which corresponds to Te sites that are located far from the Tl dopants. 
These results reveal that the anomalous enhancement in $(1/T_1T)$ is only seen for Te sites close to Tl dopants for $x$= 1.0\% in the SC compositions. 
%=================
%Before discussing the physical origin of the enhancement in $(1/T_1T)$ for Te sites close to Tl dopants in the superconducting compositions, we first emphasize how anomalous this behavior is when considered alongside the low temperature upturn in the resistivity. (INSERT TEXT HERE). 
%=================
It is especially unusual that the upturn in $(1/T_1T)$  is observed for the more \emph{heavily} doped composition ($x$=1.0\%), whereas typically for doped semiconductors one anticipates that $(1/T_1T)=const.$ for more heavily doped compositions close to the normal metallic state\cite{Meintjes}.
We note that in some cases of slightly doped semiconductors\cite{Kobayashi,Ikehata,Maeda} a Curie-Weiss-like temperature dependence of  $(1/T_1T)$ was observed at low fields due to the magnetization of the impurity states, but they are suppressed by high fields. 
Thus, the simultaneous low-temperature upturn in both the resistivity and $(1/T_1T)$ even for heavily doped compositions and the lack of suppression of the upturn by the application of a field of 12 T are the unique features in Pb$_{1-x}$Tl$_{x}$Te.
This differs from the case of the simple dilute ``{\it spin}''  Kondo effect with a low Kondo temperature. 
Furthermore, the increase in  $(1/T_1T)$ should be attributed to the local anomaly from the presence of Tl dopants because it is seen only at Te sites close to Tl dopants, and hence it also differs from the usual $q\ne 0$ magnetic excitation pictures.  
These features put Pb$_{1-x}$Tl$_{x}$Te in a distinct new class of doped semiconductors.
%It is quite unusual that the upturn in $(1/T_1T)$ is seen at heavier-doped  $x$=1.0 \%  than $x$=0.35 \%,  since in the usual doped semiconductor the $(1/T_1T)=const.$ appears in heavier-doped concentrations close to in the normal metallic state\cite{Meintjes}.
%Moreover, such a simultaneous upturn in both the resistivity and $1/T_1T$ is highly anomalous, putting Pb$_{1-x}$Tl$_{x}$Te in a distinct new class of doped semiconductors. 

The observation of a logarithmic upturn in the resistivity below 10 K for $x > x_c$ is reminiscent  of the Kondo effect\cite{Kondo}, but apparently in the absence of magnetic impurities\cite{Andronik,Matsushita}. 
It was previously suggested that this behavior could be accounted for by a {\it ``charge''} Kondo effect\cite{Varma,Taraphder,Costi,Dzero,Yanagisawa,Matsuura},  in which fluctuations between the two degenerate charge states of  the Tl impurities (i.e. 2$e^-$(Tl$^{1+}$)  and 0$e^-$(Tl$^{3+}$) for $x > x_c$) are screened by the free carriers,  analogous with the screening of fluctuations between two degenerate spin states in the conventional ``{\it spin}'' Kondo effect. 
Our experimental results further support this hypothesis. 
First, our data provide microscopic evidence that  (a) the static spin susceptibility $\chi_0$ deduced from $\Delta K$ shows no apparent $T$ dependence in the $T$ range 1.8 K$<T<50$ K, and (b) that the upturn in $(1/T_1T)$ upon cooling below 10 K is not suppressed even by the application of a strong magnetic field of $\sim$12 T, both of which are consistent with  the absence of spin degrees of freedom. 
%In the present case, it has been theoretically accounted for by {\it ``charge''} Kondo effect\cite{Varma,Taraphder,Costi,Dzero,Yanagisawa,Matsuura} where two nearly degenerate charge states of  2$e^-$(Tl$^{1+}$)  and 0$e^-$(Tl$^{3+}$) are possible to form a resonating valence state,  in analogy with two degenerate spin states in conventional ``{\it spin}'' Kondo effect.
%The difference from  the  {\it spin} Kondo effect is microscopically given by the facts that the static spin susceptibility $\chi_0$ deduced from $\Delta K$ shows no apparent $T$ dependence in the  $T$ range of 1.8 K$<T<50$ K, and the upturn in $(1/T_1T)$ upon cooling below 10 K was not suppressed even in the application of the strong magnetic field $\sim$12 Tesla, which are consistent with  the absence of spin-degree of freedom. 
%Therefore, the results may be attributed to some coherent $6s$-electron pair hopping in the resonating valence state. 
% has no spin degree of freedom, the enhancement of $(1/T_1T)$ below $T_{\rm min}^{\rm NMR}$ is not attributed to some increase of {\it dynamical} spin susceptibility $\chi''(q,f)$ as in the magnetic compounds. 
%Accordingly, it is possible that the significant increase in $(1/T_1T)$s below 10 K for $x > x_c$ is associated with the {\it ``charge''} Kondo effect as well, originating from the resonating valence state between 2$e^-$(Tl$^{1+}$) and 0$e^-$(Tl$^{3+}$). 
Second, Miyake {\it et al.} have pointed out recently that  the observed enhancement in $(1/T_1T)$ below $T_{\rm min}^{\rm NMR}$ can be readily understood within the charge Kondo picture as the result of the  $T$ dependence of the electron-pair hopping (EPH) interaction $J_{\rm ph}$ between $6s$ pair electrons on the Tl dopants and the conduction electrons\cite{Matsuura,MiyakeNMR}. 
That is, there exists a process contributing to ($1/T_{1}T$) at the Te site, which is proportional to the EPH interaction $J_{\rm ph}$. 
%Recently, Miyake {\it et al.} have pointed out that  the origin of enhancement in $(1/T_1T)$ below $T_{\rm min}^{\rm NMR}$ is explained by $T$ dependence of the electron-pair hopping (EPH) interaction $J_{\rm ph}$ between $6s$ pair electrons on the Tl dopants and the conduction electrons on the basis of the charge Kondo effect\cite{MiyakeNMR}. 
%Namely, there exists a process contributing to ($1/T_{1}T$) at Te site, which is proportional to the EPH interaction $J_{\rm ph}$. 
This EPH interaction is transformed to the spin exchange interaction $J_{\perp}$ by a particle-hole transformation for the down spin component of both the $6s$  and conduction electrons. 
The renormalized $J_{\perp}(T)$ (or $J_{\rm ph}$) by the Kondo effect exhibits a logarithmic $T$ dependence at $T\gsim T_{\rm K}$ (Kondo temperature)\cite{Anderson}, like $J_{\perp}(T)=J_{\perp}^{\infty}/[1+ J_{\perp}^{\infty}N_{\rm 0}\log(T/D)]\approx - (J_{\perp}^{\infty})^{2}N_{\rm 0}\log(T/D)$, 
where $J_{\perp}^{\infty}$ and $D$ are the bare exchange interaction and the half-bandwidth of the conduction band, respectively. 
In addition,  this scenario also reproduces no apparent $T$ dependence of the static susceptibility observed by the Knight shift.
%The results are simultaneously provide microscopic evidence for Kondo behavior while also definitively establishing the absence of spin degrees of freedom. 
These observations are consistently accounted for by the dynamical charge fluctuations in the absence of paramagnetism, which are anticipated by  a charge Kondo scenario associated with the Tl dopants. 
%Both are anticipated in a charge Kondo scenario.  
%The renormalized $J_{\perp}(T)$ (or $J_{\rm ph}$) by the Kondo effect has the logarithmic $T$ dependence at $T\gsim T_{\rm K}$(Kondo temperature)\cite{Anderson}, like 
%$ J_{\perp}(T)=J_{\perp}^{\infty}/[1+ J_{\perp}^{\infty}N_{\rm F}\log(T/D)]\approx - (J_{\perp}^{\infty})^{2}N_{\rm F}\log(T/D)$, 
%where $J_{\perp}^{\infty}$, $N_{\rm F}$, and $D$ is the bar exchange interaction, the DOS of conduction electrons at the Fermi level, and the Fermi energy of conduction electrons, respectively. 
%\[
%1/T_1T \propto U_{ph}
%\]
%where $U_{ph}$ is pair hopping interaction 
%The heavily doped semiconductors shows 
%1/T1T exhibits divergent increase at  T < TK (???½`10K) 
%through T-dependence of Uph by charge Kondo effect
%In this context, the present NMR evidence demonstrates that the resonating valence state between 2$e^-$(Tl$^{1+}$)  and 0$e^-$(Tl$^{3+}$) results in the development of coherent electron-pair hopping upon cooling below 10 K, increasing $(1/T_1T)$ around the Tl dopants for the SC sample. 
%This event may be in analogy with the increase in $(1/T_1T)$ just below $T_c$ due to the formation of Cooper-pairing, that is known as the coherence effect for BCS superconductors\cite{Hebel}. 
%This observation provides microscopic evidence for dynamical charge fluctuations in the absence of paramagnetism, and is consistent with expectations for charge Kondo behavior associated with the Tl dopant ions. 
In contrast, such anomalies in $(1/T_1T)$ were not detected for the non-SC parent ($x$=0) and for the composition close to $x_c$ ($x \sim$ 0.35\%). 
Our observations therefore provide a microscopic insight into the previously suggested connection between dynamical valence fluctuations and the occurrence of superconductivity in this material system\cite{Taraphder,Costi,Dzero,Yanagisawa,Matsuura}.
%In contrast, such anomalies in $(1/T_1T)$ were not detected in the non-SC samples ($x$=0 and 0.35 \%),  suggesting an intimate connection between dynamical valence fluctuations and the occurrence of superconductivity. 
%Thus, we deduce that the coherent hopping of  $6s$-electron pair may develop a Cooper-pairing formation to cause the superconductivity, which is suggested as the negative-$U$ scenario for the onset of superconductivity\cite{Taraphder,Costi,Dzero,Yanagisawa,Matsuura}.

Thus far we have focused on the strong spatial dependence of the Knight shift and $(1/T_1T)$ as a function of the distance to a Tl impurity, and on the $T$ dependence of these quantities for temperatures below $T_{\rm min}^{\rm NMR}\sim 10$K.  
An additional anomalous aspect of the data shown in Fig. \ref{f3} for the SC composition is the high-$T$ behavior in $1/T_1T$ $\emph{above}$ $T_{\rm min}^{\rm NMR}$. 
That is,  we find that $1/T_1T$ $\emph{increases}$ up to $T\sim$50 K, but only for the  Te sites at A$_i$ ($i$=1,2,3) in the vicinity of the Tl dopants, while remaining constant as a function of temperature for the Te sites (A$_4$) far from the Tl dopants. 
The precise microscopic origin of this effect is not yet known. 
However, since this is a local anomaly enhanced around the Tl dopants, it should certainly be attributed to these impurities. 
An intriguing possibility is that this effect is related to the crossover from a coherent resonating valence state below $T_{\rm min}^{\rm NMR}$ to incoherent excited valence states at high $T$\cite{Matsuura}. 
Such a crossover  would strongly affect the temperature dependence of the relaxation process. 
Further measurements directly probing the spin dynamics on the Tl sites might provide further insight into this behavior.

Finally, we comment on the relation of our observations to recent band structure calculations that aim to elucidate the nature of Tl impurities in PbTe. 
These calculations have revealed deep and resonant states formed from the hybridization of Tl impurities with coordinating Te ions\cite{Ahmad,Xiong}. 
Such states preserve much of the character of the original Tl wavefunctions, exhibiting similar behavior as  virtual bound states. 
For small Tl concentrations ($x < x_c$), the Fermi energy is higher in the valence band than the resonant impurity levels\cite{Xiong}. 
Hole doping moves the Fermi energy deeper into the valence band, and earlier heat capacity data for double-doped systems indicate that the Fermi energy eventually resides in the resonant impurity levels for superconducting compositions\cite{Erickson,Nemov}. 
Significantly, our measurements provide direct microscopic evidence for the anticipated spatial variation in DOS associated with such resonant impurity states.
% (illustrated schematically in Fig. 4). 
Moreover, we have shown that the nuclear spin dynamics are very different when the Fermi energy lies in the resonant impurity levels (i.e. for $x \sim$ 1.0\%) relative to when it lies farther away from them ($x \sim$ 0.35\%). 
The charge Kondo behavior suggested in the resistivity, and here in our NMR data, imply that the resonant impurity states at the Fermi energy are characterized by a negative effective U, consistent with the strongly Tl character of the associated orbitals\cite{Ahmad,Xiong} and expectations for this valence-skipping element\cite{Varma}.

In summary, systematic measurements of the $^{125}$Te NMR Knight shift and $(1/T_1T)$ on Pb$_{1-x}$Tl$_{x}$Te have revealed that the Tl dopants induce spatially inhomogeneous electronic states around the Tl dopants.  
In the SC sample with $x$=1.0\%,  a remarkable increase in $(1/T_1T)$ is observed upon cooling below 10 K for the Te sites close to the Tl dopants. 
%, is from the same origin as that in the resistivity due the development of the resonating valence state between 2$e^-$(Tl$^{1+}$) and 0$e^-$(Tl$^{3+}$), i.e. the charge Kondo effect. 
The upturn in  the resistivity is corroborated by  a microscopic probe of $(1/T_1T)$. 
A simultaneous upturn in these values is not suppressed by the application of a high field. 
Such unique features can be understood consistently in terms of coherent charge fluctuations anticipated in the charge Kondo effect arising from the Tl impurities.
%This behavior provides an additional experimental fact for a scenario based on the charge Kondo effect arising from the Tl impurities, obtained here from a direct microscopic probe. 
In contrast, such an anomaly was not detected in non-SC samples with $x$=0 and 0.35\% ($x \sim x_c$). 
Although the link between resonant impurity levels, valence fluctuations, and charge Kondo behavior has been discussed previously for this material\cite{Varma,Taraphder,Costi,Dzero,Matsuura,Yanagisawa}, these specific measurements are important in terms of providing the first microscopic insight into such a scenario. 
Although it has been established from a theoretical perspective that valence fluctuations can provide an effective pairing interaction\cite{Varma,Taraphder,Costi,Dzero,Matsuura,Yanagisawa}, it remains to be seen to what extent this is helpful for understanding the high critical temperature in this remarkable superconductor. 

%This is because the hybridization of the wave functions between widely-spread $6s$ electrons and conduction electrons becomes more significant in the SC sample, and hence yields the coherence of the resonating valence state between 2$e^-$(Tl$^{1+}$) and 0$e^-$(Tl$^{3+}$) over the sample. 
%As a result, we suggest that the coherent hopping of $6s$-electrons pair may develop the Cooper-pairing formation, which  is called the ``negative-$U$'' scenario, to cause the superconductivity below $T_c\sim$1 K.
%\cite{Taraphder,Costi,Dzero,Yanagisawa,Matsuura}.

%\section*{Acknowledgements}

{\footnotesize 
We thank H. Matsuura for his valuable comments. 
This work was supported by Izumi Science and Technology Foundation, Toyota Riken Scolar, and JSPS KAKENHI (Grant Nos. 26400356, 26610102, 16H04013, 25400369, and 17K05555). 
The work at Stanford University was supported by AFOSR Grant No. FA9550-09-1-0583.
}

%::::::::::::::::bibliography::::::::::::::::::::::::::::::::::::::::::::::::
%:::::::::::::::::::::::::::::::::::::::::::::::::::::::

\end{document}